# A Validation of the Proposed Component-Based Development Process

M. Rizwan Jameel Qureshi
Dept. of Computer Science, COMSATS Institute of Information Technology, Lahore
anriz@hotmail.com
Ph # (92-42-5431602) Cell # (03334492203)
M. E. Sandhu
National College of Business Administration & Economics
40 E/I Gulberg III Lahore, Pakistan

## Abstract

*Component-based development (CBD) is a name, with which software development professionals are quite familiar. There are several models which have been proposed for CBD in last few years. They contain good features but there are some improvement possibilities in them. The objective of this paper is to propose a process for CBD and to evaluate the effects of quality parameters on reusability. The validations of the proposed CBD model provide positive indication for software (SW) industry that it can be successfully implemented for CBD projects.*

**Key words**: Process, CBSE, CBD, domain engineering, reusability, library, quality

## 1. Introduction

A software model is the most significant arrangement in software development arena. From a simple web page to a complex multi-tier corporate system, a suitable process model is the essential requirement to ensure the reliability and success of the product [1,2,3]. Software industry is practicing various classic models as well as modern smart architectural solutions to meet the current rapidly changing user and system requirements [4,5,6]. A number of papers have been written about CBD model in the last few years [7,8,9].

The objective of this paper is to propose a process for CBD. The proposed process uses library at analysis phase instead of design phase to develop complex systems [2,10]. It also evaluates effects of quality parameters on reusability by presenting an equation based on the responses of the survey. Section 2 proposes the new Process. Section 3 describes significant features of the proposed process. Section 4 describes validation of the proposed process using a case study and a survey from sixteen software companies. Section 4 also presents the effects of quality parameters on reusability.

## 2. The Proposed Process for CBD

The proposed model for CBD has following phases.

- Project Planning
- Analysis, component & Risk Management
- Customization, Composition & Testing
- Customer Evaluation

Figure 1 shows the main phases of proposed process for CBD.

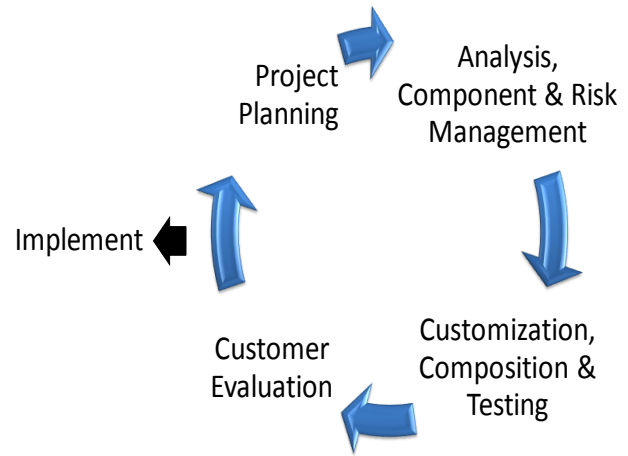

**Figure 1- Diagrammatic View of the Proposed Process**

A customer is communicated at the start of the project to gather basic requirements. Initial use cases are developed at this stage to prepare project specification or proposal document during the *planning* phase. Project specification or proposal document is composed of feasibility and risk assessments that are performed to prepare a cost benefits analysis (CBA) sheet. CBA sheet helps customers to decide whether the SW project is feasible for their organization or not.

*Analysis* phase is initiated if the customer approves the proposal. After the approval of the proposal in the planning phase, detailed requirements are gathered during the '*Analysis, Component and Risk Management*' phase. Here software engineering team identifies and selects reusable components from in-house library of components. Risks about new and existing components are also evaluated and managed. The properties, behavior and relationships among components are identified as well.

The reusable components are customized according to the requirements of the new system and tested during 'Customization, Composition *& Testing*' phase. The new components are designed, developed and tested on a unit basis. Integration and system tests of the new and reusable components are performed as well. The customer is asked to evaluate and verify SW before implement.

## 3. Significant Features of the Proposed Process

Component-based software engineering (CBSE) main activities are component identification, qualification, adaptation and composition [2]. Component qualification activity makes sure that the selected component implements the required functionality, composes seamlessly into the architecture of new application, and possesses the quality attributes [11].

'*Analysis, Component and Risk Management*' phase of the proposed process helps to identify and select reusable components rather than reinventing the wheel. This improves productivity and efficiency of software engineers [12,13]. Deployment of library at analysis phase is a promising feature of the proposed process that helps to identify and qualify components. Risk management handles the risks regarding components.

'Customization, Composition & Testing' phase deals with adaptation, integration and testing of reusable components.

## 4. Validation of the Proposed Process for CBD

Initial validation of the proposed CBD process model is conducted by two methods that are:
- A case study in a software company;
- A survey from sixteen software development companies using questionnaire.

### 4.1 Validation Using Case Study

A case study was conducted to evaluate the proposed process in a software organization. The successful implementation of this case study presented a positive sign for the suitability of the proposed process. Case study was conducted for a software company which implemented the proposed process to develop and implement leasing software for leading company of the UK. The detail to implement the proposed process is as follows.

**'Project Planning' Phase**
The analysts met client to gather basic requirements. A feasibility report was made and approved by the client. Nine month's time was estimated to customize the leasing software for the client. According to the plan two and half months would have to be spent on complete analysis, one and half month on core designing and five months on coding, testing and maintenance.

**'Analysis, Component & Risk Management' Phase**
Two analysts and one quality assurance leader were responsible to prepare functional specification after gathering complete requirements. The PM conducted an impact/gap analysis meeting with all team members. Main objectives of the impact analysis meeting were to:
- identify the reusable components.
- measure the impact of the changes to customize the reusable components for the client.

A library of previously developed components was available at a software company where case study was conducted. The library contained complete information of all parts of the SW. It was used during the impact analysis to search the components which were reusable.

**'Customization, Composition & Testing' Phase**
One and half month were spent on design according to the schedule prepared during the planning phase. Detailed use cases, ERD, normalization and main user interfaces were developed for the user stories (that were different from client 1) by database developers. The Project Manager decided for two weeks deadline to complete a build after completion of three months of core designing. Five months were planned to customize, compose and test the reusable components.

The programmers integrated their components into the application after conduction of unit, integration and system tests within two weeks time. The quality assurance engineers again conducted unit, integration and system tests for all the components. The programmers were requested to fix the bugs if code crashed during testing. This process also continued for rest of the coding duration. Two workable demos were shown to the client during customization of three releases of the leasing application.

**'Customer Evaluation' Phase**
The SW was successfully coded, tested and deployed within the planned duration. A contract was made with the client company to provide maintenance of the SW.

The results of the case study show that the proposed process is implementable for CBD.

**4.2 Validation Using Survey**
A survey involving sixteen software development organizations was conducted to evaluate the proposed Process. The people who filled the forms had more than six years experience in software development. Questionnaire technique was used to gather the data. Twenty eight professionals were selected to fill the questionnaire forms. It was divided into three main sections. Each section consisted of different questions. The sections were:

- Suitability of proposed system development life cycle (SDLC) phases for CBD.
- Suitability of library of in-house components on the proposed SDLC phases for CBD.
- Effect on reusability with respect to interoperability, complexity, efficiency, reliability, upgradeability, time saving, cost and quality for proposed CBD Process.

**Effects of Proposed Process Phases for CBD Projects**
Tables 1 to 2 are based on evaluations. The parameter evaluated in Table 1 is as follows.
The parameter assessed in Table 1 is as follows.

A- Importance of Analysis, Component & Risk Management phase for CBD projects.

| Weight → | % of 1 | % of 2 | % of 3 | % of 4 |
|---|---|---|---|---|
| Parameter ↓ | | | | |
| A | 7 | 3 | 35 | 60 |

**Table 1- Importance of 'Analysis, Component & Risk Management' Phase for CBD projects**

The weight values in the questionnaire range from 1 to 4. Four reflects 'Analysis, Component & Risk Management' phase is highly suitable for CBD with respect to the parameter and one means not suitable.

The parameters mentioned in Table 2 are as follows.
- B- Suitability of 'Analysis, Component & Risk Management' phase of CBD projects.
- C- Suitability of 'Customization, Composition & Testing' phase for CBD projects.
- D- Suitability of 'Customer Evaluation' phase for CBD projects.

| Weight → | % of 0 | % of 1 |
|---|---|---|
| Parameters ↓ | | |
| B | 46 | 54 |
| C | 35 | 65 |
| D | 35 | 65 |

**Table 2- Suitability of Proposed SDLC Phases for CBD Projects**

The weight values in the questionnaire are from 0 to 1. One means the proposed SDLC phases are highly suitable for CBD with respect to the parameters and zero means not suitable.

**Results of Tables 1 to 2**
It can be concluded from Tables 1 to 2 that respondents highly supported the suitability of proposed SDLC phases for CBD projects.

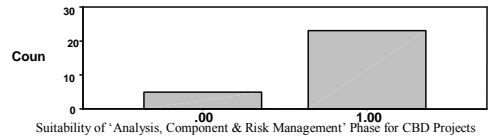

**Figure2- Suitability of 'Analysis Component Selection & Risk Management' Phase for CBD Projects**
Figure 2 shows suitability of the analysis, component and risk management phase for CBD projects. It shows that most of the SW developers are of views that the 'analysis, component and risk management' phase is extremely important for development of CBD projects.

**Suitability of Library/Repository of In-house Components on the Proposed Process for CBD**
The parameter evaluated in Table 3 is as follows.

A-Library needs at 'Analysis Component Selection & Risk Management' Phase

| Weight → | % of 0 | % of 1 |
|---|---|---|
| Parameter ↓ | | |
| A | 46 | 54 |

**Table 3- Library Needs at 'Analysis, Component & Risk Management' Phase**

The weight values in the questionnaire are from 0 to 1. One means library is highly needed for 'Analysis, Selection & Risk Management' phase with respect to the parameter and zero means not needed.

Parameters assessed in Table 4 are as follows.

    B- Importance of Library for CBD projects.

    C- Library helps to classify reusable components.

    D- Library makes easier to search reusable components.

    E- Library facilitates modification of reusable components.

    F- Library helps to test reusable components.

    G- Library facilitates implementation of reusable components.

    H- Library makes it easier to manage versions of reusable components.

    I- Library helps to maintain up to date and consistent documentation.

| Weight → / Parameters ↓ | % of 1 | % of 2 | % of 3 | % of 4 |
|---|---|---|---|---|
| B | 3 | 25 | 35 | 28 |
| C |  | 14 | 50 | 35 |
| D |  | 7 | 65 | 28 |
| E |  | 40 | 35 | 25 |
| F | 7 | 32 | 35 | 25 |
| G | 3 | 10 | 65 | 21 |
| H |  | 21 | 25 | 60 |
| I | 3 | 17 | 35 | 42 |

**Table 4-Effect of Library on the Proposed Process for CBD**

The weight values in the questionnaire range from 1 to 4. Four indicates very high suitability of library on the proposed process with respect to the parameters and one means not suitable.

**Results of Table 4**

The results of Table 4 show that library is highly suitable on the proposed process for CBD projects.

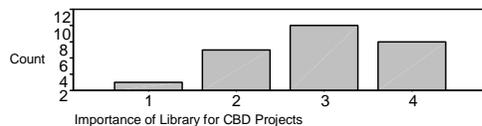

**Figure 3- Importance of Library for CBD Projects**

Figure 3 shows importance of library for CBD projects. It shows that most of SW developers are of the view that library significantly improves development of CBD projects.

**Effect on Reusability With respect to Quality Parameters for CBD Projects**

Likert scale value from 1 to 4 evaluates the effects of quality parameters on reusability for CBD projects.

- 1 means low effect on reusability
- 2 means average effect on reusability
- 3 means high effect on reusability
- 4 means very high effect on reusability

Univariate analysis [14] is performed to conclude the results.

**Results of Table 5**

It can be concluded from results of Table 5 (after references) that respondents highly supported improvement in reusability because of interoperability, complexity, efficiency, reliability, upgradeability, cost saving and quality for CBD projects.

Table 5 provides R square value. It shows that variation in quality is 0.869 because of defined variables. It is expected that variables have strong contribution if value of R square $> 0.7$ and weak if value of R square $< 0.3$ [15]. The variables in this model have very significant contribution because value of R square is 0.869 for the defined model. Therefore it can be concluded that the model is highly reliable.

*Reusability = - 0.163 + 0.176 interoperability + 0.472 complexity + 0.003354 efficiency - .00868 reliability + 0.111 upgradeability + 0.359 time saving – 0.190 cost + 0.172 quality*

The equation indicates effect of independent variables (Interoperability, complexity, efficiency, reliability, upgradeability, time saving, cost and quality) on the dependent variable (reusability). The adjustment factor in equation is $-0.163$ which calculates the effect of quality parameters on reusability by SPSS statistical software package itself [15].

The defined model suggests that if a unit:
- increases interoperability it increases 0.176 units in reusability;
- increases complexity it increases 0.472 units in reusability;
- increases efficiency it increases 0.003354 units in reusability;

- increases reliability it decreases 0.00868 units in reusability;
- increases time saving it increases 0.359 units in reusability;
- increases cost it decreases 0.190 units in reusability;

These results show quality metrics (interoperability, complexity, efficiency, reliability, upgradeability, time saving, cost saving and quality) have very significant effect on reusability. This shows that the proposed process is reasonably useful for software companies using CBD.

## 5. Conclusion

A new process has been presented for the component-based development. A case study has been conducted for a software organization to validate the proposed process. A survey involving sixteen software development organizations has also been carried out to evaluate the proposed process. An equation has been developed from the data to estimate the effect of quality parameters on reusability. The results of the equation show that the proposed model is highly reliable and quality factors have significant effect on reusability. The proposed process provides strong support for reusability, interoperability, upgradeability, less complexity, time saving, cost saving, reliability and also improved quality. The validation of the proposed process for CBD provides an indication for its usability for software industry.

## 6. References

[1] Sebastian Tyrrell, "The Many Dimensions of the Software Process," Crossroads ACM Press, vol. 6, no. 4, pp. 22-26, 2004.

[2] Roger S. Pressman, "Software Engineering". McGraw Hill, USA, 2005.

[3] A. Schmietendorf, E. Dimitrov, R. R. Dumke, "Process Models for the software development and performance engineering tasks," in Proc. 3$^{rd}$ Int. workshop on Software and performance, Rome, Italy, 2002, pp. 211–218.

[4] I. Crnkovic, M. Larsoon, "Building Reliable Component-Based Software Systems", Artech House, 1$^{st}$ Ed., 2002.

[5] Dogru, A.H., Tanik, M.M., "A process model for component-oriented software engineering," IEEE Software, vol. 20, no. 2, pp.34-41, March-April 2003.

[6] M.R.V. Chaudron, http://www.win.tue.nl/~mchaudro/cbse2005/01_IntroCBD_Concepts.pdf. Visited 6 August 2008.

[7] de Almeida, E.S., Alvaro, A., Lucredio, D., Garcia, V.C., de Lemos Meira, S.R., "A survey on software reuse processes," IEEE Int. Conf. on Information Reuse and Integration, Aug. 2005, pp. 66-71.

[8] Gerald Kotonya, Ian Sommerville, Steve Hall, "Towards A Classification Model for Component-Based Software Engineering Research," in Proc. 29$^{th}$ Conf. on EUROMICRO, 2003, pp. 43.

[9] Crnkovic,I. Larsson, S. Chaudron, M., "Component-based development process and component lifecycle," in Proc. 27th Int. Conf. Information Technology Interfaces, June 20-23, 2005, pp. 591-596.

[10] Nasib S. Gill, "Reusability issues in component-based development," ACM SIGSOFT Software Engineering, vol. 28, no. 4, 2003, pp. 4-4.

[11] Luiz Fernando Capretz, Miriam A.M. Capretz, Dahai Li, "Component-Based Software Process," in Proc. 7th Int. Conf. Object-Oriented Information Systems, August 27-29, 2001, pp. 523-529.

[12] M. R. J. Qureshi, S. A. Hussain, "A Reusable Software Component -Based Development Process Model," International Journal of Advances in Engineering Software, Feb. 2008, vol. 39, no. 2, pp. 88-94.

[13] M. R. J. Qureshi, S. A. Hayat, "The Artifacts of Component-Based Development," Science International Journal Lahore, 2007, vol. 19, no. 3, pp. 187-192.

[14] East Carolina University, http://core.ecu.edu/psyc/wuenschk/SPSS/SPSS-Lessons.htm. Visited August 1, 2008.

[15] M. Hanif, M. Ahmad, A. M. Ahmad. "Biostatistics for Health Students". An ISOSS Publication, Pakistan, 2004.

**Tests of Between-Subjects Effects**

Dependent Variable: Reusability

| Source | Type III Sum of Squares | df | Mean Square | F | Sig. |
|---|---|---|---|---|---|
| Corrected Model | 240.168a | 8 | 30.021 | 112.791 | .000 |
| Intercept | .506 | 1 | .506 | 1.902 | .170 |
| INTER | 1.686 | 1 | 1.686 | 6.333 | .013 |
| COMP | 8.212 | 1 | 8.212 | 30.855 | .000 |
| EFF | .112 | 1 | .112 | .420 | .518 |
| REL | .178 | 1 | .178 | .668 | .415 |
| UPGRD | .585 | 1 | .585 | 2.198 | .141 |
| TS | 3.267 | 1 | 3.267 | 12.276 | .001 |
| CS | 1.176 | 1 | 1.176 | 4.418 | .038 |
| QULT | .857 | 1 | .857 | 3.222 | .075 |
| Error | 33.803 | 127 | .266 | | |
| Total | 1510.000 | 136 | | | |
| Corrected Total | 273.971 | 135 | | | |

a. R Squared = .877 (Adjusted R Squared = .869)

**Table 5 General linear model for Univariate Analysis**

Here:
df = Degree of freedom
F = It is a statistical test in which the test statistic has an F-distribution if the null hypothesis is true. F-test is used to reach at a decision regarding mean differences.
P= Significance of the defined model

**Parameter Estimates**

Dependent Variable: Reusability

| Parameter | B | Std. Error | t | Sig. |
|---|---|---|---|---|
| Intercept | -.163 | .118 | -1.379 | .170 |
| INTER | .176 | .070 | 2.517 | .013 |
| COMP | .472 | .085 | 5.555 | .000 |
| EFF | 3.354E-02 | .052 | .648 | .518 |
| REL | -8.68E-02 | .106 | -.817 | .415 |
| UPGRD | .111 | .075 | 1.483 | .141 |
| TS | .359 | .102 | 3.504 | .001 |
| CS | -.190 | .091 | -2.102 | .038 |
| QULT | .172 | .096 | 1.795 | .075 |

**Table 6 Parameter Estimated**